\documentclass[aps,pra,superscriptaddress]{revtex4}
\usepackage{mathrsfs}
\usepackage{amsfonts}
\usepackage{graphicx}
\usepackage{amssymb}
\usepackage{amsmath}
\usepackage[retainorgcmds]{IEEEtrantools}
\begin{document}
\def\ran{\rangle}
\def\lan{\langle}
\def\cl{\centerline}
\def\bd{\begin{description}}
\def\be{\begin{enumerate}}
\def\ben{\begin{equation}}
\def\benn{\begin{equation*}}
\def\een{\end{equation}}
\def\eenn{\end{equation*}}
\def\benr{\begin{eqnarray}}
\def\eenr{\end{eqnarray}}
\def\benrr{\begin{eqnarray*}}
\def\eenrr{\end{eqnarray*}}
\def\ed{\end{description}}
\def\ee{\end{enumerate}}
\def\al{\alpha}
\def\b{\beta}
\def\bR{\bar\R}
\def\bc{\begin{center}}
\def\ec{\end{center}}
\def\dg{\dagger}
\def\d{\dot}
\def\D{\Delta}
\def\del{\delta}
\def\ep{\epsilon}
\def\g{\gamma}
\def\G{\Gamma}
\def\h{\hat}
\def\iny{\infty}
\def\La{\Longrightarrow}
\def\la{\lambda}
\def\m{\mu}
\def\n{\nu}
\def\noi{\noindent}
\def\Om{\Omega}
\def\om{\omega}
\def\p{\psi}
\def\pr{\prime}
\def\r{\ref}
\def\R{{\bf R}}
\def\ra{\rightarrow}
\def\up{\uparrow}
\def\dn{\downarrow}
\def\lr{\leftrightarrow}
\def\s{\sum_{i=1}^n}
\def\si{\sigma}
\def\Si{\Sigma}
\def\t{\tau}
\def\th{\theta}
\def\Th{\Theta}

\def\vep{\varepsilon}
\def\vp{\varphi}
\def\pa{\partial}
\def\un{\underline}
\def\ov{\overline}
\def\fr{\frac}
\def\sq{\sqrt}
\def\ot{\otimes}
\def\tf{\textbf}
\def\WW{\begin{stack}{\circle \\ W}\end{stack}}
\def\ww{\begin{stack}{\circle \\ w}\end{stack}}
\def\st{\stackrel}
\def\Ra{\Rightarrow}
\def\R{{\mathbb R}}
\def\mf{\mathbf }
\def\bi{\begin{itemize}}
\def\ei{\end{itemize}}
\def\i{\item}
\def\bt{\begin{tabular}}
\def\et{\end{tabular}}
\def\lf{\leftarrow}
\def\nn{\nonumber}
\def\va{\vartheta}
\def\wh{\widehat}
\def\vs{\vspace}
\def\Lam{\Lambda}
\def\sm{\setminus}
\def\ba{\begin{array}}
\def\ea{\end{array}}
\def\bd{\begin{description}}
\def\ed{\end{description}}
\def\lan{\langle}
\def\ran{\rangle}
\def\l{\label}
\def\mb{\mathbb}
\def\ti{\times}
\def\mc{\mathcal}
\def\v{\vec}
\large

\preprint{}
\title{ Invariance of quantum correlations under local channel for a bipartite quantum state.}

\date{\today}
\author{Ali Saif M. Hassan}
\email{alisaif73@gmail.com}
 \affiliation{Department of Physics, University of Amran, Amran, Yemen}

\author{Pramod S. Joag}
\email{pramod@physics.unipune.ac.in}
\affiliation{Department of Physics, University of Pune, Pune, India-411007.}

\date{\today}
\begin{abstract}
We show that the quantum discord in a bipartite quantum state is invariant under the action of a local quantum channel if and only if the channel is invertible. In particular, quantum discord is invariant under a local unitary channel.

\noi PACS numbers: 03.65.Ud;03.65.Db;03.65.Yz
 \end{abstract}

\maketitle

\emph{Introduction} : The characterization and quantification of quantum correlations in multipartite quantum systems is a fundamental problem facing quantum information theory \cite{niel,horo,olli,hv,lf}. Two principal approaches regarding quantum correlations consist of entanglement separability paradigm and the quantum- versus classical paradigm \cite{horo,olli,hv,lf,modi,lew,chen,rudo,guo,horo1,stom,xi,wang,wang1,augu,wu,gess,siew,hou,qi,lf1,hou1}. In the last decade the later approach generated intense research activity, as it was found that quantum correlations involved in separable states play a crucial role in some applications \cite{knil,dat1,dat2}. In the quantum vs. classical paradigm, the quantum correlations form the part of total correlations which arise over and above the classical correlations implied by a multipartite quantum state. Such quantum correlations are characterized by quantum discord (QD)\cite{olli,hv}, measurement induced non-locality (MiN) \cite{lf}, quantum deficit  \cite{opp} etc. These quantum correlations can be a resource for a number of quantum information applications \cite{modi,lew,daki,asma,roa}. In this context it becomes important to understand the dynamics of these quantum correlations under local noise (local quantum operations).\\

The basic problem we are concerned with is the relation of the local quantum operation (a local CPTP map or local quantum channel) on a part of a bipartite system in a given state with the resulting change in quantum correlations as the state changes under such local operations. In particular, we are concerned with the possibility of changing quantum discord under such local quantum operations. For two qubit case it is shown that the qubit channel that preserves commutativity is either unital i.e. mapping maximal mixed state to maximal mixed state, or a completely decohering channel that nullifies QD in that state \cite{str}. For $m\times 3$ system, it is shown that a channel acting on a second subsystem cannot create QD for zero QD states if and only if it is either a completely decohering channel or an isotropic channel \cite{hu}.It was further proved that a channel $\Lambda$ transforms a zero QD state to zero QD state if and only if $\Lambda$ preserves commutativity \cite{hu,yu}. The exact forms of unital channel for qubit system and that for the completely decohering channel for any system are also obtained \cite{yj}.\\

In this paper we obtain condition on the local quantum operation (channel) operating on any bipartite system to preserve quantum correlations, namely the quantum discord. This result is completely general and applies to any bipartite state. The results obtained tackle an interesting problem ; that of characterizing the class of operations that do not change measures of nonclassical correlations such as quantum discord  \cite{cla,gal}.\\

\emph{Main results}:
Our idea is to express the mutual information $I(\rho)$ and the classical correlations $C_{A,B}(\rho)$ (defined below) in a bipartite state $\rho$ in terms of the relative entropy involving two states $\rho$ and $\si$ given by $$S(\rho||\si)=-S(\rho)-tr(\rho \log{\si})$$ where $S(\rho)= -tr(\rho \log \rho)$ is the von-Neumann entropy of the state $\rho.$ We then seek condition on the local quantum operation so as to preserve $I(\rho)$ and $C_{A,B}(\rho)$ expressed in terms of the relative entropy. We need the following

\emph{Theorem 1}  (Petz \cite{petz,hayd}) : For states $\rho$ and $\si$ and a local quantum operation $T$ $$S(\rho||\si)=S(T\rho||T\si)$$ if and only if there exists a local quantum operation $\h{T}$ such that $$\h{T}T\rho = \rho \; ;\;\h{T}T\si = \si .$$

We now state and prove the main results.

\emph{Lemma 1 }: Let $\rho_{AB}$ be a bipartite quantum state, $\Lambda_{a}\;:\;\rho_{A} \longmapsto \Lambda_{a}\rho_{A}$ ; $\Lambda_{b}\;:\;\rho_{B} \longmapsto \Lambda_{b}\rho_{B}$ be local quantum operations and $\rho_{A,B}=tr_{B,A}\rho_{AB}.$ Then $$I((\Lambda_{a}\ot \Lambda_{b})\rho_{AB})\;=\;I(\rho_{AB})$$ if and only if there exist quantum operations $\Lambda^{*}_{a,b}\;:\;\rho_{A,B} \longmapsto \Lambda^{*}_{a,b}\rho_{A,B}$ such that $$(\Lambda^{*}_{a}\ot \Lambda^{*}_{b})(\Lambda_{a}\ot \Lambda_{b})\rho_{AB}=\rho_{AB},$$ and $$(\Lambda^{*}_{a}\ot \Lambda^{*}_{b})(\Lambda_{a}\ot \Lambda_{b})\rho_A\otimes \rho_B = \rho_A\otimes \rho_B.$$ Here $I(\rho)$ is the quantum mutual information in the state $\rho_{AB}.$

\emph{Proof} : We use the definition of the quantum mutual information via relative entropy, namely, $$I(\rho_{AB})=S(\rho_{AB}||\rho_{A}\ot \rho_{B}).$$  Then the result is immediate from Theorem 1. We just have to replace $\rho$ by $\rho_{AB},$ $\si$ by $\rho_{A}\ot \rho_{B}$ and the quantum operation $T$ by $\Lambda_{a}\ot \Lambda_{b}.$ $\blacksquare$

Next we obtain the conditions on the map $(\Lambda_{a}\ot \Lambda_{b})$ which leave the classical correlations in a bipartite quantum state $\rho_{AB}$ unchanged. Classical correlations in a bipartite quantum state $\rho_{AB},$ with a POVM measurement defined by the set of positive operators $\{A^{\dg}_{i}A_{i}\}$ $(\{B^{\dg}_{i}B_{i}\})$ carried out on part A(B) can be expressed as  \cite{hv}
\ben \l{e1}
C_{A,B}(\rho_{AB}) = \max_{\{A^{\dg}_{i}A_{i}\},\{B^{\dg}_{i}B_{i}\}}[S(\rho_{B,A})- \sum_{i}p_i S(\rho^{i}_{B,A})]      \\
  = \max_{\{A^{\dg}_{i}A_{i}\},\{B^{\dg}_{i}B_{i}\}}\sum_{i}p_i S(\rho^{i}_{B,A}||\rho_{B,A})       \\
\een
where $\rho^{i}_{B,A}=tr_{A,B}\rho^{i}$ is the reduced density operator of the density operator $\rho^{i}$ prepared by the POVM measurement $\{A^{\dg}_{i}A_{i}\},\{B^{\dg}_{i}B_{i}\}$ after $i$th outcome.

\emph{Lemma 2} : Let $\rho_{AB}$ be a bipartite quantum state and $\Lambda_{a,b}\;:\;\rho_{A,B} \longmapsto \Lambda_{a,b}\rho_{A,B}$ be the local quantum operations where $\rho_{A,B}=tr_{B,A}\rho_{AB}.$ Then $$C_{A,B}((\Lambda_{a}\ot \Lambda_{b})\rho_{AB})=C_{A,B}(\rho_{AB})$$ if and only if there exist local quantum operations $\Lambda^{*}_{a,b}\;:\;\rho_{A,B} \longmapsto \Lambda^{*}_{a,b}\rho_{A,B}$ such that $$(\Lambda^{*}_{a}\ot \Lambda^{*}_{b})(\Lambda_{a}\ot \Lambda_{b})\rho_{AB}=\rho_{AB},$$ and $$(\Lambda^{*}_{a}\ot \Lambda^{*}_{b})(\Lambda_{a}\ot \Lambda_{b})\rho_A\otimes \rho_B = \rho_A\otimes \rho_B.$$ Here $C_{A,B}(\rho)$ denote the classical correlations in the state $\rho$ as defined in Eq.(\r{e1}).

\emph{Proof} : Note that $\rho_{AB}$ and hence $\rho_A$ and $\rho_B$ are arbitrary density operators so that the domain and the range of $\Lambda$ and $\Lambda^{*}$ cover the corresponding set of all density operators. In particular,  for the states $\rho^{i}_{B,A}$  occurring in Eq. {(\r{e1})
$ \Lambda^{*}_{b,a}\Lambda_{b,a}(\rho^{i}_{B,A})=\rho^{i}_{B,A}$ and similarly for $\rho_{B,A}$.
The result follows because by theorem 1 each term in the sum defining $C_{A,B}(\rho_{AB})$ (Eq.(\r{e1})) is invariant under the action of the local channels $\Lambda_{a}$ and $\Lambda_{b}$ as defined in the statement of the lemma. Hence the maxima over the corresponding POVMs $\{A^{\dg}_{i}A_{i}\},\{B^{\dg}_{i}B_{i}\}$ are also invariant.$\blacksquare$

\emph{Theorem 2} : Let $\rho_{AB}$ be a bipartite quantum state and $\Lambda_{a,b}\;:\;\rho_{A,B} \longmapsto \Lambda_{a,b}\rho_{A,B}$ be the local quantum operations where $\rho_{A,B}=tr_{B,A}\rho_{AB}.$ Then $$D_{A,B}((\Lambda_{a}\ot \Lambda_{b})\rho_{AB})=D_{A,B}(\rho_{AB})$$ if and only if there exist local quantum operations $\Lambda^{*}_{a,b}\;:\;\rho_{A,B} \longmapsto \Lambda^{*}_{a,b}\rho_{A,B}$ such that $$(\Lambda^{*}_{a}\ot \Lambda^{*}_{b})(\Lambda_{a}\ot \Lambda_{b})\rho_{AB}=\rho_{AB}.$$ and $$(\Lambda^{*}_{a}\ot \Lambda^{*}_{b})(\Lambda_{a}\ot \Lambda_{b})\rho_A\otimes \rho_B = \rho_A\otimes \rho_B.$$  Here $D_{A,B}(\rho_{AB})= I(\rho_{AB}) - C_{A,B}(\rho_{AB})$ denote the quantum discord in the state $\rho_{AB}.$  We also assume that the domain and the range of all the local maps cover the corresponding set of all density operators.

\emph{Proof} : The \emph{if} part follows immediately from lemma 1 and lemma 2. For the \emph{only if} part, we have to cater to the possibility that a non-invertible local channel can preserve discord by lowering the classical corelation (a local operation cannot increase the classical correlation see \cite{hv}) and mutual information by equal amount, or, in other words, by lowering the classical correlation alone, without changing the quantum correlation. Thus we must show that there is no non-invertible local channel which preserves quantum discord by changing the classical correlation alone for \emph{all} states (although a particular non-invertible local map may preserve discord of a particular state by reducing $I(\rho_{AB})$ and $C_{A,B}(\rho_{AB})$ by the same amount). To do this, we partition all the non-invertible local channels into two (mutually exclusive) classes, namely the commutativity preserving channels \cite{hu} and those which are not commutativity preserving. The non-commutativity-preserving channels are shown to create discord in zero discord (classical quantum) states of a bipartite system \cite{hu}. Hence these local channels do not preserve discord in all bipartite states. The commutativity preserving channels are shown to have one of the two forms, namely completely decohering channel and the isotropic channel \cite{hu,yj}. A completely decohering channel nullifies the quantum discord in any bipartite state. So it cannot preserve quantum discord in all discordant states. An isotropic local channel which is of the form \cite{hu} $\Lambda^{iso}(\rho)=p \Gamma(\rho) + (1-p) \frac{I}{d},$ where $\Gamma$ is any linear channel that preserves the eigenvalues of $\rho$ and $d$ the dimension of Hilbert space, decreases quantum correlation. To see that, act by $I\otimes \Lambda^{iso}(\rho)$ on maximally quantum correlated pure state $|\psi\ran=\frac{1}{\sqrt{2}}(|00\ran+|11\ran)$ which gives a mixed state which has less quantum correlation. So, the isotropic local channel can not preserve quantum correlation for all states. Then, the commutativity-preserving channel does not preserve quantum correlation for all states (although it preserves discord for a zero discord state). This covers all the non-invertible local channels and completes the proof. $\blacksquare$

The most useful invariant local quantum channels are the local unitary operators $\rho \longmapsto U\rho U^{\dagger}$ and the anti-unitary operator $\Theta = UK$, $U$ unitary and $K$ the conjugation operator $\rho \longmapsto UK\rho KU^{\dagger}$ \cite{saku}. However, the anti-unitary operator, which corresponds to the time-reversal, is unphysical if applied only to a part of a quantum system. Therefore, the anti-unitary local operators must act on both parts of the system. Thus we have shown, in particular, that the local unitary channel preserves quantum discord. Finally, it will be interesting to search for the characteristics of local channels which increase discord in (possibly some class of) bipartite quantum states with non-zero discord.

\emph{Acknowledgments} : This work was supported by the BCUD grant RG-13. ASMH acknowledges University of Pune for hospitality during his visit when this work was carried out.

\end{document}